\documentclass{article}
\usepackage{arxiv}
\usepackage[utf8]{inputenc} 
\usepackage[T1]{fontenc}    
\usepackage{hyperref}       
\usepackage{url}            
\usepackage{breakurl}
\usepackage{booktabs}       
\usepackage{amsfonts}       
\usepackage{nicefrac}       
\usepackage{microtype}      
\usepackage{lipsum}
\usepackage{graphicx}
\usepackage{float}
\usepackage{pifont}
\newcommand{\cmark}{\ding{51}}%
\newcommand{\xmark}{\ding{55}}%

\usepackage{ulem}
\newcommand*\samethanks[1][\value{footnote}]{\footnotemark[#1]}
\graphicspath{ {Images/} }
\graphicspath{ {./Images/} }

\title{IoT meets COVID-19: Status, Challenges, and Opportunities}

\author{
 Mohsin Kamal\thanks{Authors contributed equally.}\\National University of Computer\\and Emerging Sciences, Pakistan \\ mohsin.kamal@nu.edu.pk
   \And
 Abdulah Aljohani\samethanks\\ King Abdulaziz University\\ Saudi Arabia\\ ajaljohani@kau.edu.sa
  \And
 Eisa Alanazi\samethanks\\Umm Al-Qura University\\Saudi Arabia\\ eaanazi@uqu.edu.sa}

\usepackage{ulem}

\begin{document}
\maketitle
\begin{abstract}
   	Due to the global pandemic of COVID-19, there is an urgent need to utilize existing technologies to their full potential. Internet of Things (IoT) is regarded as one of the most trending technologies with a great potential in fighting against the coronavirus outbreak. The IoT comprises of a scarce network in which the IoT devices sense the environment and send useful data on the Internet. In this paper, we examine the current status of IoT applications related to COVID-19, identify their deployment and operational challenges, and suggest possible opportunities to further contain the pandemic. Furthermore, we perform analysis for implementing IoT in which internal and external factors are discussed.  
\end{abstract}
\section{Introduction}
The Internet of Things (IoT) consists of a complex network of smart devices that frequently exchange data over the Internet \cite{srinivasan2019review}. It has renovated the actual world objects into clever virtual objects. The goal of IoT is to unite everything in our world under a mutual arrangement, helping the users in not only controlling the objects around them but also keeping them up-to-date about the state of the things \cite{panarello2018blockchain}. IoT devices sense the environment and send the acquired data to the Internet cloud without the requirement of human-to-human or human-to-machine interaction. IoT has become an integral part in today's modern era of communication where tens of millions of devices are connected via IoT and the number is growing rapidly \cite{al2020survey}.

\begin{figure}[htb]
\centering
\includegraphics[width=0.8\textwidth]{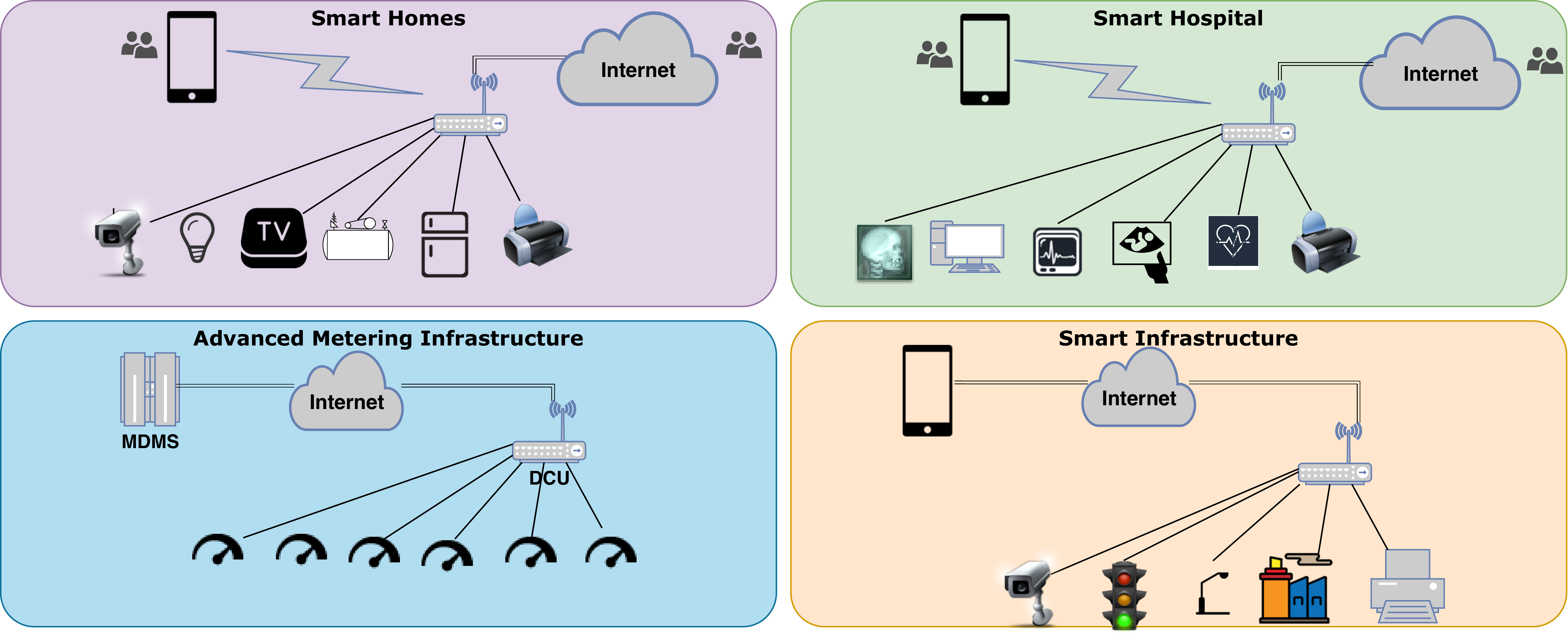}
\caption{Several IoT applications including: large-scale deployment of home, hospital, utility metering, and infrastructure.}
\label{fig:IoT_cocept}
\end{figure}

IoT has the potential to play a vital role in various fields of life, such as health systems \cite{saranya2019iot}, autonomous vehicles \cite{minovski2020modeling}, home and industrial automation \cite{aheleroff2020iot}, intelligent transportation \cite{mohsinv2v}, smart grids \cite{yin2019toward}..etc. An overview of IoT is given in Figure \ref{fig:IoT_cocept} in which IoT is deployed in health sector, homes, power sector and various infrastructures. Sensors obtain data of related information from the environment and use the Internet cloud as a medium of delivering information to the relevant body or organization \cite{kamal2018light}. The core concept behind IoT is the realization of multiple devices communicating with each other seamlessly. This has the promise of better utilization of available resources, reduction of cost, and minimizing manual interaction. 
As the 2019 coronavirus disease (COVID-19) continues to spread across the globe, it is inevitable to discuss and articulate the IoT potential during pandemics. As of June 21, 2020, the number of COVID-19 confirmed cases has exceeded ~8.5 million \cite{covidPK} with 3.7\% mortality rate \cite{mehta2020covid}. Researchers from different fields continue to investigate and generate diverse solutions which could help in combating the COVID-19 \cite{mavrikou2020development}. 

IoT comes up with the ingredients needed to help the countries in minimizing the effect of COVID-19. IoT has a wide range of applications which would be effective to make sure that all the guidelines of safety and precautions provided by health officials are followed \cite{pham2020artificial}. IoT has a scalable network which has the potential to deal with huge amount of data received from sensors used by number of applications to fight against COVID-19. Furthermore, the reliable IoT networks decrease the delivery time of crucial information which can help in providing timely response during the global pandemic of COVID-19 \cite{allam2020coronavirus}. The role of IoT was never needed to the extent to which it is required now because of coronavirus outbreak.

This paper discusses the effectiveness of IoT in combating the global pandemic of COVID-19. Additionally, several scenarios are examined in which IoT can help in reducing the outbreak of coronavirus. Furthermore, we have analyzed the possible challenges that the IoT-based solutions encounter in combating the coronavirus. 

The rest of paper is organized as follows. Section \ref{sec:applications} presents the important applications of IoT in the perspective of COVID-19. Challenges in implementing the IoT are described in Section \ref{sec:Challenges}. SWOT analysis is performed in Section \ref{sec:SWOT}. Solutions of tackling the challenges in deploying the IoT are presented in Section \ref{sec:solutions}.

\section{Applications of IoT to combat Covid-19} \label{sec:applications}

\begin{figure}[H]
\centering
\includegraphics[width=0.8\textwidth]{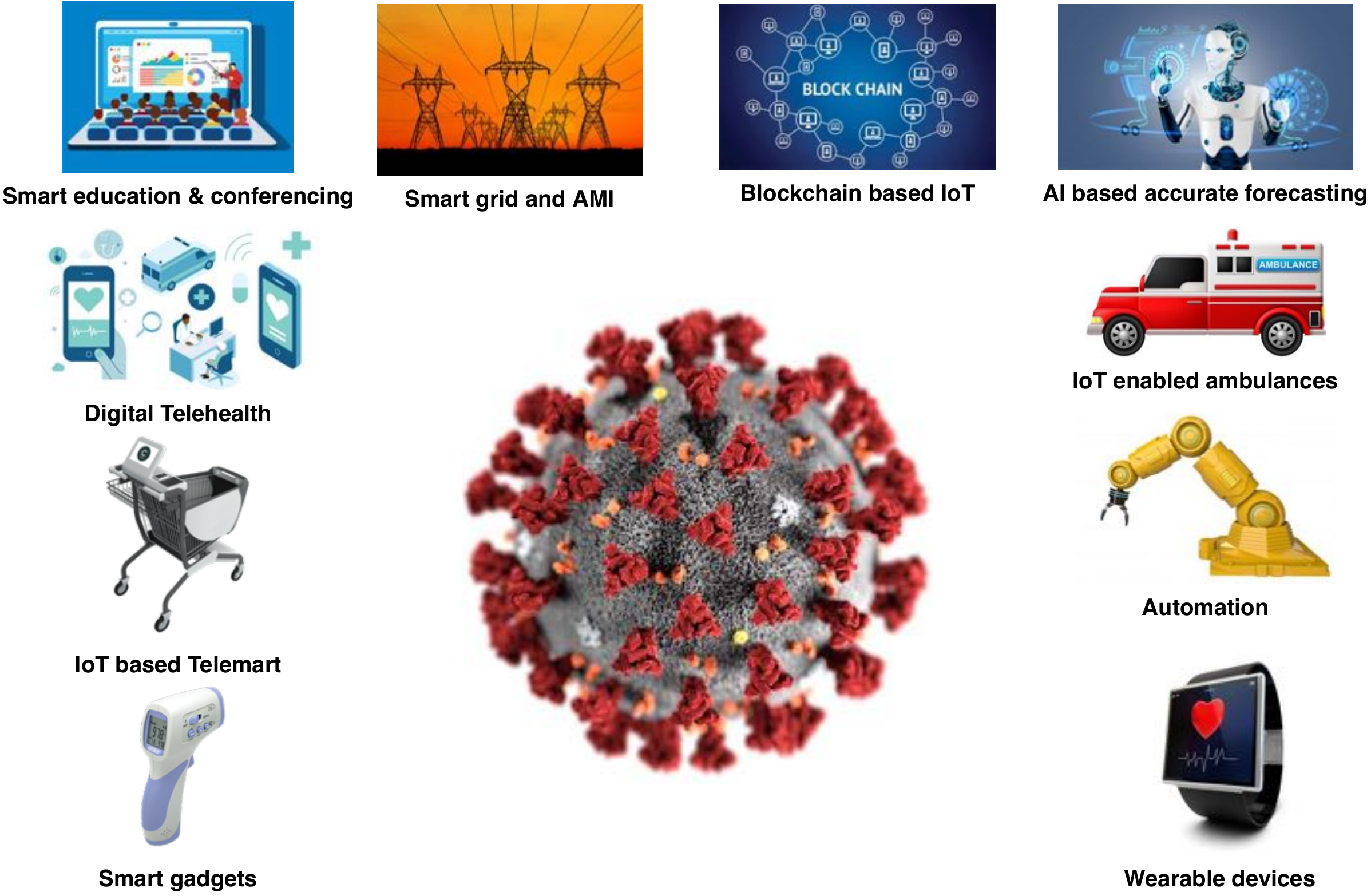}
\caption{Potential IoT applications to combat COVID-19.}
\label{fig:IoT_Applications}
\end{figure}

The seamless connections and vigorous integration with other technologies has enabled the IoT to be one of the promising technology that will change our lives \cite{hussain2020machine}. The applications of IoT in combating this global pandemic can be spread to several sectors which can play a major role in reducing the risk of coronavirus outbreak \cite{saeed2020wireless}. Figure \ref{fig:IoT_Applications} shows potential applications in which IoT technologies can be useful and effective into combating the COVID-19. The following subsections will examine the capability of the IoT in fighting against COVID-19.

\subsection{Internet of Health Things and Digital Telehealth} \label{sec:IoHT}
\begin{figure}[hbp]
\centering
\includegraphics[width=0.7\textwidth]{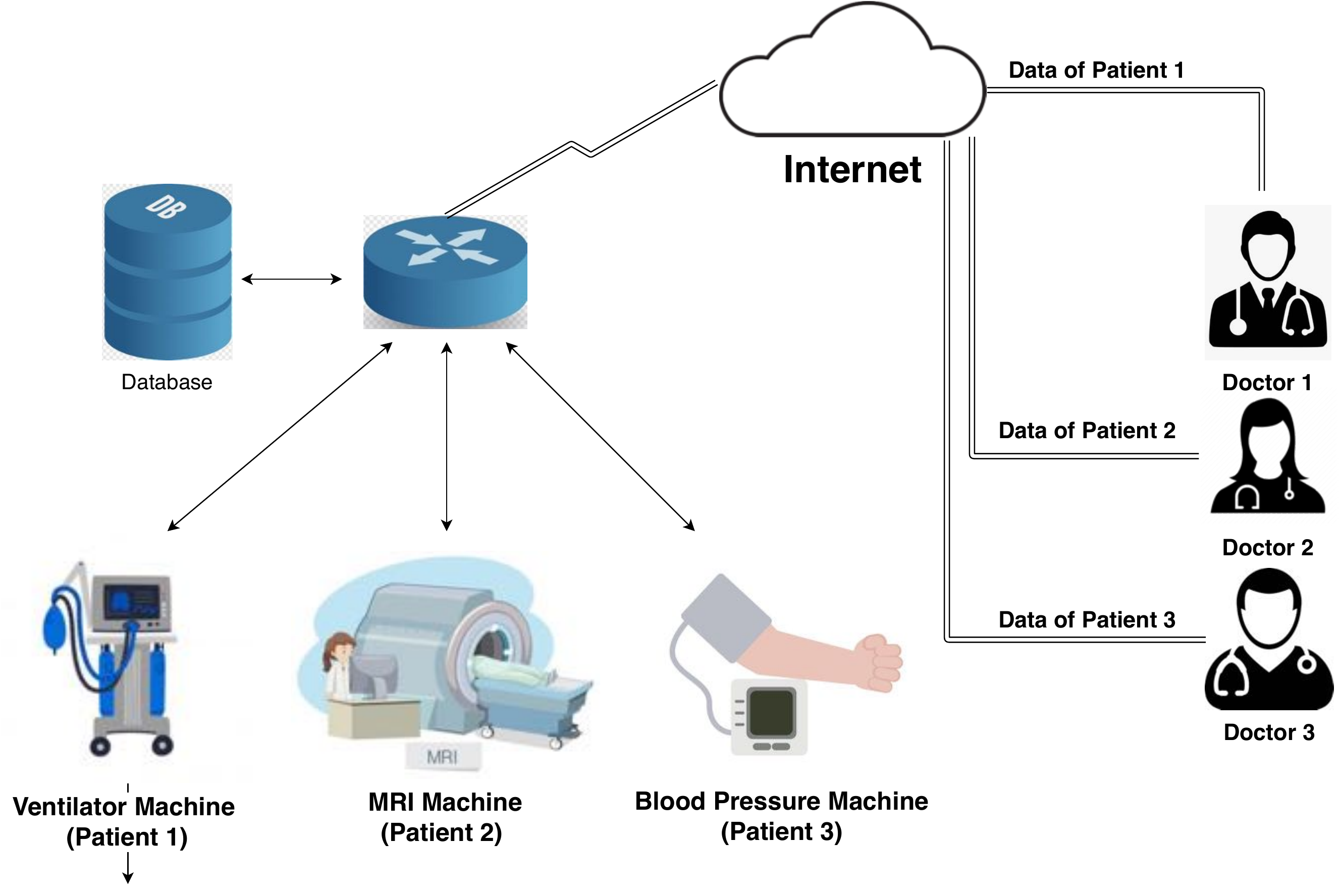}
\caption{Remote examination of medical patients by the doctors in IoT}
\label{fig:IoT_Health}
\end{figure}

Internet of Health Things (IoHT) is an extension of IoT, which aims to connect patients to health care facilities to monitor and control human body vital signs using communication infrastructure \cite{rodrigues2018enabling}. Telemedicine is getting popular in remote areas where accessibility to a quality physician is limited due to different factors \cite{telehealth123}. For example, monitoring of heart rate, electrocardiography, diabetes, and vital body signs can be remotely monitored without the physical presence of patients. An example of the remote data acquisition using IoHT system is shown in Figure \ref{fig:IoT_Health}. 
The sensors and actuators receive data from patient and send the information to the cloud using a local gateway. The doctor examines the data using any mobile or desktop application provided to them and notifies the patient or medical staff taking care of the patient about the report \cite{poppas2020telehealth}. 

Digital telehealth plays a very important role during COVID-19 outbreak. A portal is created where patients interact with the doctors and the treatment is provided remotely. The benefit of employing a secured IoHT system in COVID-19 is that the physicians do not come in direct contact with the patients, hence, avoiding the  the spread of virus \cite{telehealth123}. Many countries have started operating the digital telehealth in this time of crisis. Health Arc \cite{healthArc} provides IoT based health care devices to the patients whose data is continuously monitored by the medical staff. The data is analyzed and the suggestions and prescription are provided to the patients on their mobiles or tablets. Continuous Care \cite{continiouscare}, Health net connect \cite{healthnetconnect} and Sehetyab \cite{sehatyab} are among the leading telehealth service providers. A person with COVID-19 symptoms can use assessment tool provided on the digital platform such as "COVID-19 Gov PK mobile app" \cite{covidPK}, which is accessed by the physicians remotely. Using this tool, patients are timely guided and many precious lives can be saved. Furthermore, it also serves to reduce the number of hospitalizations, readmissions, and density of patients in hospitals, all of which help in improving the quality of life and provides timely treatment to COVID-19 patients.

%

\subsubsection{IoT enabled ambulances}
\begin{figure}[h]
\centering
\includegraphics[width=0.7\textwidth]{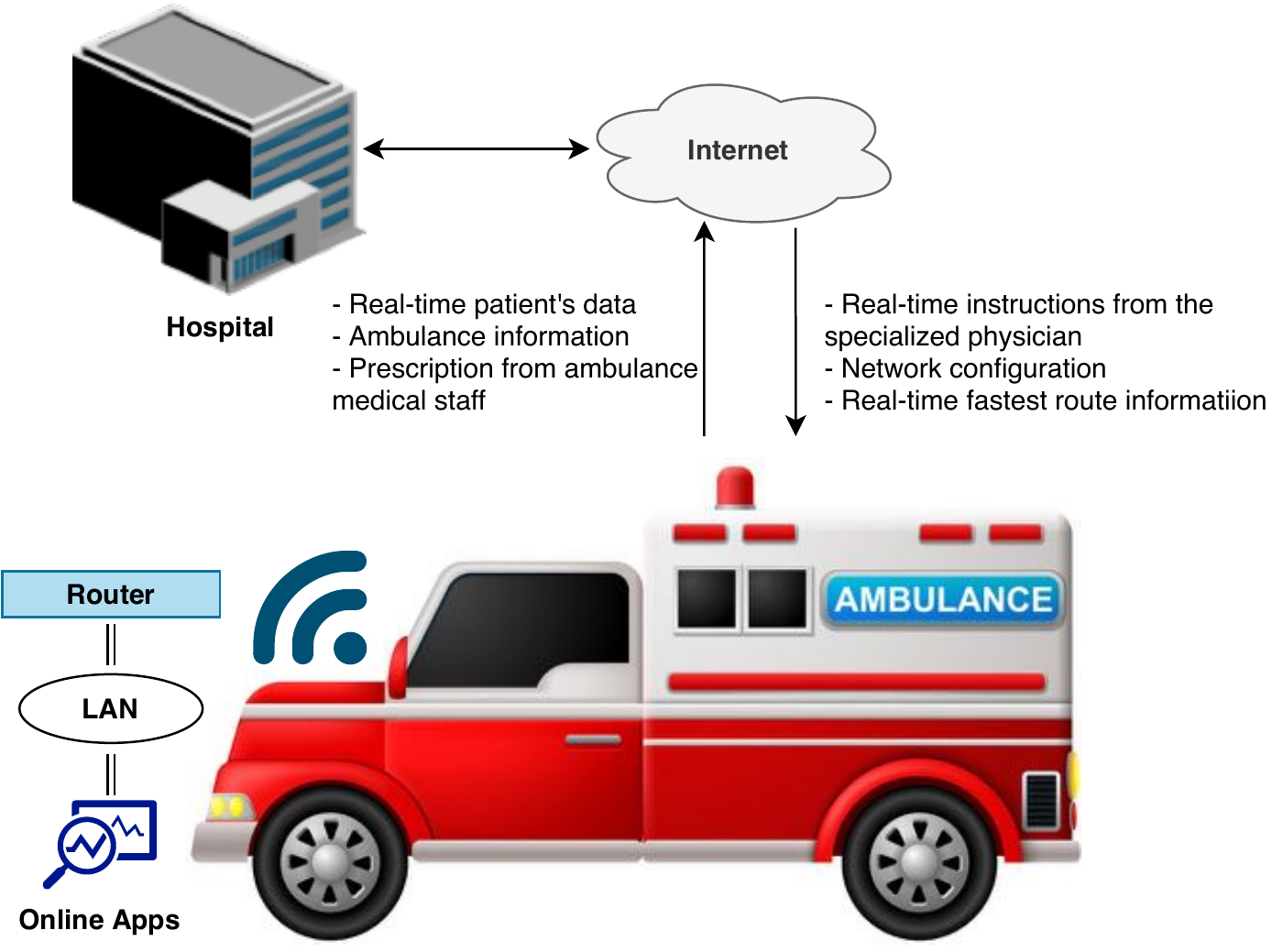}
\caption{IoT based smart ambulance system}
\label{fig:ambulance}
\end{figure}

Medical staff associated with ambulances are usually dealing with very high pressure and error prone situations \cite{park2018requirement}. During the current pandemic of COVID-19, the situations have become even more tensed and pressurized for medical staff dealing COVID-19 patients. The IoT-aided ambulances offer an effective solution in which remote medical experts suggest necessary actions to the medical staff dealing with the patient in the ambulance. This leads to the timely response and effective handling of patient. Figure \ref{fig:ambulance} shows the smart ambulance which is equipped with IoT based technology. WAS vehicles \cite{ambulance} provides smart solutions based emergency vehicles. The radio-frequency identification (RFID) based equipment is connected to wireless local area network (WLAN). The information of patient is remotely accessible by the concerned medical staff.

\subsubsection{Smart health monitoring gadgets}
IoHT-aided equipment is classified into two categories i.e., personal and clinical \cite{habibzadeh2019survey}. Personal-aided IoHT gadgets are used for self monitoring of health \cite{ghanavatinejad2019clustering}. The most common gadgets used are Apple watch \cite{applewatch} and Fitbit \cite{fitbit} etc. The user tracks the heartbeat, exercise, sleep, nutrition and weight using these gadgets. These are useful in fighting against COVID-19 as well because the rest and sleep become very important factors for the patients suffering from this disease. Sleeping well increases the immunity of the body to fight against the virus \cite{bai2020presumed}. The patient can see his reports on the portals provided by these gadget makers and provide information to the related physicians if required. IoT based wearable gadgets can help in reducing the spread of coronavirus if certain algorithms are implemented to the existing devices. The wearable devices notify in real time if:
\begin{itemize}
\item The social distancing protocol is violated,
\item any COVID-19 patient is in the locality, and
\item the area declared as danger zone by the government in the perspective of coronavirus outbreak.
\end{itemize} 

\begin{figure}
\centering
\includegraphics[width=0.35\textwidth]{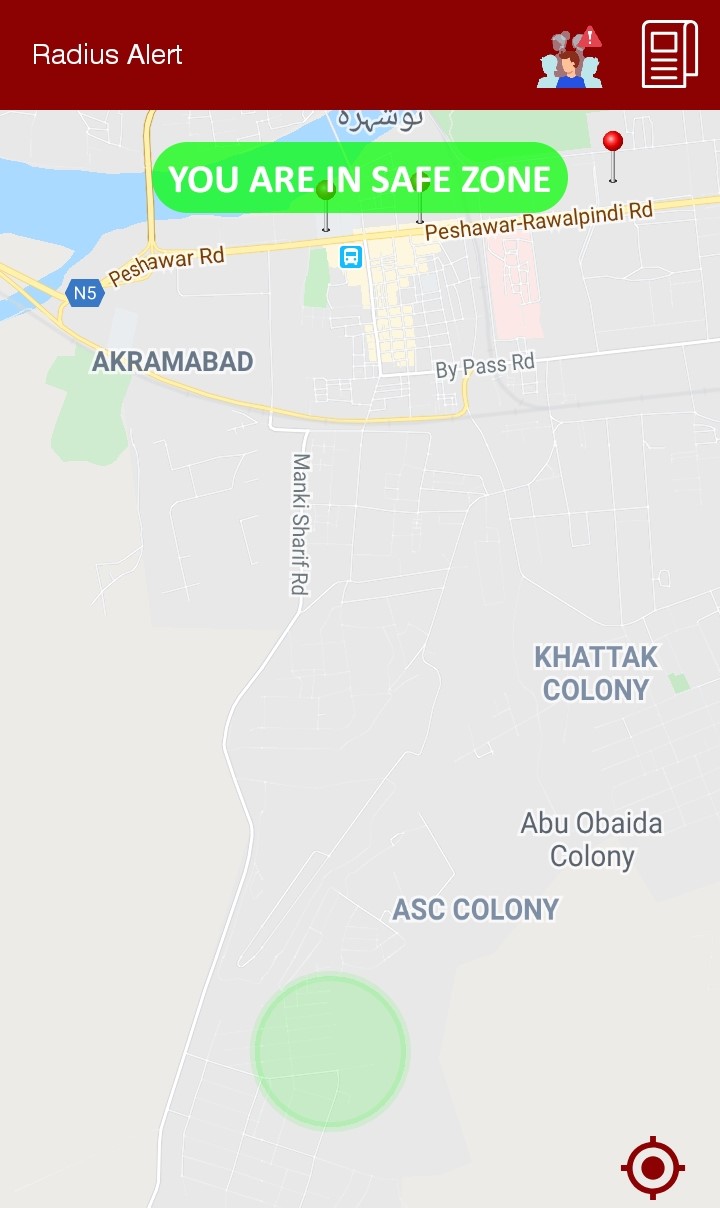}
\caption{Real-time area monitoring for Covid-19 patients}
\label{fig:safe_danger_zone}
\end{figure}

Figure \ref{fig:safe_danger_zone} shows the presence of user in safe zone \cite{covidPK}. If a user moves in the area which is danger zone with respect to COVID-19 patients, the device intimates in real time and user can maximize precautions to stay safe from coronavirus. 

Clinical IoHT includes the monitoring of person under the supervision of physician as shown in Figure \ref{fig:IoT_Health}. A list of IoT enabled clinical gadgets is presented in \cite{clinical}. The list includes devices to monitor the spread of cancer, continuous glucose monitoring, connected inhalers, asthma monitor and many more. During this global pandemic, many healthcare gadgets provide opportunities of real time remote supervision. These gadgets are smart enough to provide results which can be seen remotely by medical staff \cite{revathi2020influence}. It can have a system to activate alarm if any unforeseen situation occurs. In the perspective of COVID-19, IoT based ventilators and temperature monitors can help in providing the timely assistance to patients. The patient status can be monitored remotely if ventilators are connected to the cloud. The IoT based temperature monitoring device can help in keeping the real time record of everyone in the database. The record can be checked in later date if required \cite{mohammed2020novel}. 

\subsubsection{AI based forecasting and social distancing}
Artificial Intelligence (AI) is one of the most promising technology which is helping in bringing revolution in many fields. Introducing AI algorithms to the IoT has opened new doors in this area. AI provides opportunity of learning and extracting meaningful patterns from the data. As the data from IoT devices are aggregated in a database, it can be readily used in predicting the outbreak and effects of coronavirus and how to mitigate it \cite{pham2020artificial}. The data of COVID-19 patients helps in predicting the future behavior of virus and performing the area wise comparison of its effects. Besides, it also helps in possible match of COVID-19 symptoms in effective and quick way. The AI based treatment aids in quick recovery and monitoring of patients as well \cite{vaishya2020artificial}. The medical record of patient and the outcomes generated helps in predicting the better treatment choices based on AI and machine learning (ML) algorithms \cite{frobesAI}. BlueDot \cite{bluedot} was among the first AI power company to forecast the spread of coronavirus and identifying the global threat. They provided information on the mobility pattern of the virus along with its outbreak potential. Other AI powered companies who joined hands and worked in fighting against COVID-19 are Deargen \cite{deargen}, Insilico Medicine \cite{insilico}, SRI Biosciences and Iktos \cite{sri}, Benevolent AI \cite{bene}, DeepMind \cite{deepmind}, Nanox, Baidu, Alibaba and EndoAngel Medical Technology Company \cite{top10}. 

\begin{figure}
\centering
\includegraphics[width=0.7\textwidth]{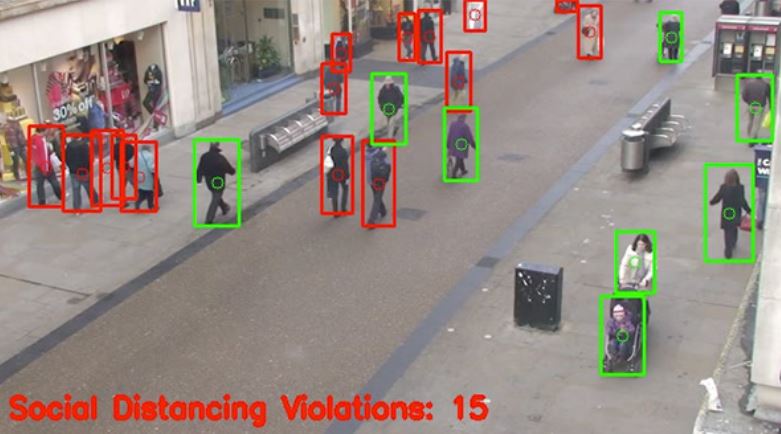}
\caption{Social distance detector using deep learning, computer vision \cite{DL}.}
\label{fig:deep_learning}
\end{figure}

Computer vision and deep learning provide real time social distancing measurement by detecting people within 2m or 6ft from each other which runs real-time on GPU and includes pre-trained model \cite{DL}. In Figure \ref{fig:deep_learning}, the red box indicates that the social distance protocol is violated while green marks safe distancing. The information is continuously sent on the Internet where the authorizing body takes action in real time if violation of social distancing is detected. This helps in minimizing the spread of coronavirus and is one of the strongest tool to help in combating the current pandemic of COVID-19. AI based emergency traffic control opens the way to the ambulances and other emergency service providers. Red Ninja \cite{redninja} is a Liverpool based company who developed an algorithm called Life First Emergency Traffic Control (LiFE) that enables paramedics to use real-time data about congestion to manipulate traffic \cite{AIforambulances}.

\subsection{Industrial Internet of Things}
Industrial Internet of Things (IIoT) is a subset of IoT which covers machine-to-machine (M2M) communication and equips industrial communication technologies with automation \cite{sisinni2018industrial}. It covers broad range of applications and provides substantial help in economic growth of countries \cite{hatzivasilis2018industrial}. In current pandemic of COVID-19, IIoT can help in sustaining the economy. It also provides support for industries to carry out its functions remotely. 
 
\subsubsection{Smart infrastructure}
The automatic human body temperature sensing machines are installed in many countries where camera is integrated with the sensor and sends real time information to the server. The system also uses AI to recognize face and matches it with the centralized database \cite{konstantakopoulos2019deep}. Use of these devices helps in tracking the patient of COVID-19. Social distancing is enforced by enabling smart infrastructure in which the environment is sensed and reports are generated in real time for the law enforcement agencies \cite{gupta2020enabling}. 
The data from sensors is constantly recorded on an online database for constant monitoring \cite{mehmood2020smart}. The content of hazardous gases or carbon content is communicated to the environmental protection agency and is updated on the server to be checked online \cite{ullo2020advances}. Researchers are working on sensing the coronavirus which can also be operated in the same manner. IBM, Microsoft, Huawie and Cisco are among the top companies who provide smart infrastructure solutions \cite{smartinfra}. 
\subsubsection{Industrial automation}
The advent of autonomous systems and their rapid evolution has played a pivotal role in designing safety systems for industries, infrastructure and transit systems \cite{javaid2020industry}. These safety systems have significantly minimized the occurrence of anomalies resulting due to human errors or environmental factors by providing early warnings and taking necessary preventive measures \cite{hafeez2020efficient}. By integrating IoT to the existing autonomous systems, the benefits and positive effects can be witnessed at large extent \cite{choo2018cryptographic}. In many countries these days, the industries have stopped its operations because of the coronavirus outbreak. Industrial automation helps in performing operations remotely. Besides, the industries monitor all the functions and hazards and the operations are performed by a single tap using smart phones.  
\subsubsection{Telemart} 
The social distancing protocols are defined by WHO as 1 meter or 3 feet \cite{WHO}. In super markets and malls, it becomes hard to maintain this distance. So the concept of telemart is introduced with the help of IoT. One of the common example is Amazon go in which the user scans a QR code at the gate of the store and starts shopping \cite{amazon_go}. After completing shopping, the user packs items and leaves. The amount is charged from Amazon account. The IoT based shopping system keeps track of all the items that are picked from the store and communicate with the sensors of the shopping cart. This is the best solution to keep social distancing by not standing up in lines to pay. 

\subsubsection{Smart Grid}
In these times of global crisis, the importance of continuous power supply has increased. All the hospitals, emergency service providers and domestic users need to have no interruptions in power. In smart grids, sensors and transducers are used to monitor and control the demand of power \cite{morello2017smart}. The important element of smart grid is Advanced Metering Infrastructure (AMI).   
The traditional power grids are upgraded with the capabilities of communication and information systems as their main functions \cite{kamal2019light}. These functions are important in analyzing, monitoring and controlling devices in the system. The goals of these upgraded grids are to save energy, reduce costs, and increase reliability and transparency \cite{ghosal2019key}. Electricity Service Providers (ESPs) are installing Advanced Metering Infrastructure (AMI) which plays a key role in the detection, localization, and prevention of any malicious activity in the network commonly known as non technical losses or theft. AMI in smart grid contains Smart Meters (SMs). These SMs through the Internet sends all the information to Data Concentrator Unit (DCU) which further forwards it to Meter Data Management System (MDMS) \cite{martins2019smart}. The computations regarding theft detection, bills generation and energy consumption etc., are performed remotely using IoT \cite{8746620}. To minimize the spread of COVID-19, the system requires this application to reduce the interaction of ESPs and the customer. Also keeping in  mind the safety of many workers who manually fix any issues that come in the metering infrastructure, AMI helps is remotely accessing the meter/consumer data and refreshing the system if any fault occurs. Besides it helps in monitoring any power cut in the locality due to which timely required action can be taken. 

AMI provides priority based services quality of service (QoS). If any network cut happens, the information is sent to the MDMS via DCU \cite{rajakumar2018power}. The area which is declared as danger zone by the government during the coronavirus outbreak has the priority among all the other areas. The recovery of fault and uninterrupted power supply is provided at the earliest.  

\subsubsection{IoT for virtual education and conferencing}
In order to reduce the spread of coronavirus, all the educational institutes have stopped its operations. The faculty is advised to manage lectures remotely. Most of the corporate sector has also advised its staff to work from home. The need of virtual classes and virtual conferencing has increased like never before. E-learning and e-conferencing require digital tools, high speed and uninterrupted Internet connectivity. The platform commonly used by the students to take online lectures and by staff members to interact with each other while being working from homes are Google meet \cite{google_meet}, Zoom \cite{zoom}, Microsoft Teams \cite{teams} and Skype \cite{skype}. IoT is playing a major role in enabling end to end communication.

\section{Challenges of IoT in the wake of COVID-19} \label{sec:Challenges}
Implementing IoT is never an easy task to perform. Furthermore, when implementing IoT for COVID-19, there are many challenges involved, few of which are described below.

\subsection{Scalability}
With the advent of digital technology, the number of IoT devices are growing exponentially \cite{gupta2017scalability}. The reason is that they are not limited to only one application but there are many applications of IoT which are in practice these days. According to a recent survey, there is a massive increase in the use of home automation appliances from 2018 to 2022 \cite{richter_2018}. The trend is represented in Table \ref{tab:home_automation}.

\begin{table}
\centering
\caption{Expected growth in domestic IoT applications from 2018 to 2022 (in million units).} \label{tab:home_automation}
\begin{tabular}{cccc}\hline
\textbf{System} &  \textbf{2018} &  \textbf{2022} &  \textbf{Growth \%} \\ \hline
Video Entertainment & 310.5m & 457.5m & 10\% \\ 
Home Monitoring \& Security & 97.7m & 244.9m & 26\% \\ 
Smart Speakers & 99.8m & 230.5m & 23\% \\ 
Lighting & 37.7m & 104.6m & 29\%  \\ 
Thermostats & 13.6m & 37.5m & 29\% \\ 
Others & 84.5m & 189.3m & 22\% \\ \hline

\end{tabular}
\end{table}

Scalability is a big challenge in implementing IoT to fight against the global pandemic of COVID-19. Large number of devices are required in IoHT alone to accurately sense the vital signs of the patients and forward those to the Internet cloud. As for now, the active cases are approximately 3.7 million worldwide. Each IoT gadget needs to have multiple sensors. Implementing IoT for this highly scaled scenario is a big challenge. The devices required are large in number and large amount of data will float around these small IoT nodes. Besides, the energy requirements are also increased due to scalability.  

\subsection{Limited spectrum and bandwidth}
As the number of IoT devices are increasing, more bandwidth is required to send all the information from sensors to the cloud. At present, most of the IoT devices use the licensed spectrum offered by the mobile operators. With the growth of these devices, the bandwidth requirements have also increased. The data faces latency which sometimes causes erroneous data transfer. For fixed IoT, operators use WiFi which becomes unreliable when the number of IoT devices are increased in its coverage area. Currently, many IoT devices use 4G/LTE networks to perform their tasks. This limited spectrum of 3G/LTE/4G will soon be not enough for large number of IoT devices. 

During the pendamic of COVID-19, timely transfer of data from IoT devices to the concerning body is of utmost importance. Errors or delay in data may cause loss of precious human lives. If the bandwidth is high, the problems of latency and low data rates can be overcome. 

\subsection{Security and privacy issues}
Due to scalability and energy limitations of IoT devices, the traditional cryptographic techniques are not feasible solutions to implement security in IoT \cite{aman2017secure}. The security solutions should be energy efficient and algorithms defined to secure the IoT network should have less computational complexities to offer end to end data protection, consumer privacy and secure authentication \cite{aman2018two}. Thus, lightweight security algorithms need to be designed in order to implement security in IoT. With the outbreak of coronavirus, the security requirements of IoT enabled networks have increased. The security concerns in implementing IoT with respect to COVID-19 are:
\begin{itemize}
\item the data which is sent from the sensors attached to the body of COVID-19 patent should be accurate,
\item the data should successfully reach the destination,
\item the data should not be forged,
\item the data should not be intercepted from the communication path, and
\item the data stored in the memory of the IoT device should not be accessible to everyone.
\end{itemize}

The security primitives should be taken considering IoT devices which have low computational capabilities. Besides being lightweighted, the required security algorithms should be accurate and must be able to keep user's trust intact \cite{kamal2018light}.

\subsection{Big data centers}
As each IoT device sends data to the cloud on some pre-defined Application Program Interface (API), the large amount of data is stored at the data centers. One of the biggest challenge in implementing IoT to combat COVID-19 is that it requires big storage centers where all the required data can be stored without overloading.  

\section{Strengths, weaknesses, opportunities and threats analysis} \label{sec:SWOT}

Strengths, Weaknesses, Opportunities and Threats (SWOT) analysis for IoT is shown in Table \ref{tab:SWOT}. The internal factors are comprised of strengths and weaknesses which are limited to the organizations or researchers who want to implement IoT in any area. The internal factors can be changed with time. Opportunities and threats are considered as external factors which depend on the market and can not be changed \cite{swot}.

\begin{table}[h]
\centering
\caption{SWOT analysis of IoT in the perspective of global pandemic} \label{tab:SWOT}
\begin{tabular}{c c}\hline
\multicolumn{2}{c}{\textbf{Internal Factors}} \\ \hline
\textbf{Strengths} & \textbf{Weaknesses} \\ 
Accuracy of data & High processing server/fusion centers are required \\ 
On time treatment & Scalability of IoT devices \\ 
Timely diagnosis & Huge data centers and data aggregation \\ 
Information of safety measure & Security and privacy preservation \\ 
High demand of IoT based systems & High bandwidth requirements \\ 
Accurate forecasting & Limited spectral resources \\ \hline
\multicolumn{2}{c}{\textbf{External Factors}} \\ \hline
\textbf{Opportunities} & \textbf{Threats} \\ 
Creation of awareness about the requirement of IoT & Compatibility of devices \\
Creation of jobs & Use of unlicensed bands \\ 
Towards mmWave communication for higher bandwidths & \\ 
Software defined radios & \\ 
Cooperative communication & \\ \hline

\end{tabular}

\end{table}

\subsection{Strengths}
Considering COVID-19 as test case, accuracy of data in IoT is one of the strengths in implementing it. The sensors take real time data from the environment and sends it to the cloud. This results in helping the patients to get on time treatment which can save many lives. If anyone has symptoms of COVID-19 and needs to consult physician then IoT helps in providing platform of telehealth in which a person can take advice of physician without visiting the hospital or clinic. This refers to the timely diagnosis of COVID-19. IoT can help in spreading the awareness related to the information and safety measures to take preemptive measures against coronavirus. Due to the importance of IoT to combat against the current global crisis of coronavirus, there is a high demand of IoT based systems. Integrating AI with IoT can help in better forecasting of future needs to fight against COVID-19. 
\subsection{Weaknesses}
The shortcomings and weaknesses can not be ignored while considering the implementation of IoT in order to combat against this virus. Due to the requirements of large number of IoT devices and scalability, the data processing units should have high processing power. The data centers should be more to keep record of patients and related information. The whole IoT network should be highly secured and the security algorithms should be designed in such a way that complexity is kept as low as possible. As many devices will be sending data frequently to the cloud, the requirements of high bandwidth can not be ignored. The mechanism should be designed where limited spectrum should be efficiently used. This can be done by frequency planning and reuse mechanism.

\subsection{Opportunities}
The opportunities are huge by implementing IoT to combat this global crisis of COVID-19. Because of the paradigm shift towards the use of digital technology and smart phones, awareness about using IoT applications specifically for combating coronavirus out break is not a difficult task to perform. Besides, the IoT industry can help in providing the jobs in local markets and effectively take its part in boosting the economy of any country. The use of Millimeter wave (mmWave) and 5G have not yet come into play for IoT networks which provides large bandwidth and high data rate. The implementation of IoT can bend the tech giants toward the use of these large bandwidth mmWave which operates between 3 GHz to 300 GHz \cite{rappaport2013millimeter}. This will open new doors in many areas of wireless communication networks. Currently, software defined radios, cognitive radio networks and cooperative communication can be applied in existing IoT networks to efficiently use the spectrum by sensing the empty spaces in licensed bands and use them for its operations. 
\subsection{Threats}
The threats as external factors are few comparatively. Currently, IoT devices are compatible with the manufacturer of the same vendor. There is a dire need of compatibility to develop competition among the vendors. This will increase the quality of IoT operations and the applications will evolve with time. Besides, the range of unlicensed bands are very less. Most of the communication in IoT either uses cellular network or 2.4 GHz of Industrial, Scientific, and Medical (ISM) frequency band which may cause interference if proper planning is not performed.

\section{Solution to the challenges in combating COVID-19} \label{sec:solutions}
The challenges involved in implementing the scalable IoT networks are undeniable but solutions to these challenges are present in the literature which can help in successfully deploying the IoT networks. Some of the prominent solutions are presented in the following subsections. 

\subsection{Lightweight security algorithms}
Due to scalability, most of the IoT devices are small in size and easily accessible. Measures must be taken to ensure that the data is protected and is efficiently received at the destination. In most cases, IoT nodes are not physically protected, so data security and provenance serve as the backbone for implementing IoT networks. Data can be easily forged if the proper security primitives are not used. Security primitives include specific attack detection, channel state masking, intrusion detection, localization, and data provenance. Provenance is finding the source of the data. A single change to the data can cause major problems. For example, medical health reports of COVID-19 patient generated by IoT devices sent to doctors and interruption in power of smart grids can cause major problems during the global pandemic of COVID-19. Due to the energy limitation of IoT devices, traditional cryptographic techniques are not viable solutions. Energy-efficient security primitives that take up less memory space having less computational complexities are the key building blocks for end-to-end content protection, user authentication, and consumer confidentiality in the IoT world. Light-weight security algorithms are mostly based on simple encryption techniques. 

Various metrics like Angle of Arrival, Time of arrival, phasor information, and Received Signal Strength Indicators (RSSI) can be used to develop lightweight security algorithms for IoT. Link fingerprints are generated from communicating IoT devices. These link fingerprints are encoded with symmetric key and the resultant is sent to the server where it computes Pearosn correlation coefficient using the link fingerprints of connected IoT devices. The computation of Pearson correlation coefficient is very simple technique yet very accurate to detect any adversary in the IoT network \cite{kamal2018light}.  Some of the notable lightweight security algorithms present in literature are summarized in Table \ref{tab:comp_protocols}. 

\begin{table}
\centering
\caption{Lightweight security algorithms in literature to combat various attacks in IoT network.} \label{tab:comp_protocols}
\begin{tabular}{ccccc}\hline
Security Requirements &  Gope et al. \cite{gope2019lightweight} & Dong et al. \cite{dong2015detecting} & Ali et al. \cite{ali2014securing} & Kamal et al. \cite{kamal2018light}  \\ \hline
MITM attack & \cmark & \cmark & \cmark & \cmark   \\
Jamming & \cmark & \xmark & \cmark & \cmark   \\ 
Data Tempering & \xmark & \xmark  & \cmark & \cmark   \\
Replay attack & \cmark & \xmark & \xmark & \xmark   \\ 
Location Proximity & \xmark & \xmark & \xmark & \cmark   \\ 
Data Provenance & \xmark & \xmark & \cmark & \cmark   \\ \hline

\end{tabular}
 
\end{table}

\subsubsection{Blockchain for connected healthcare units and privacy preservation}
\begin{figure}[h]
\centering
\includegraphics[width=0.7\textwidth]{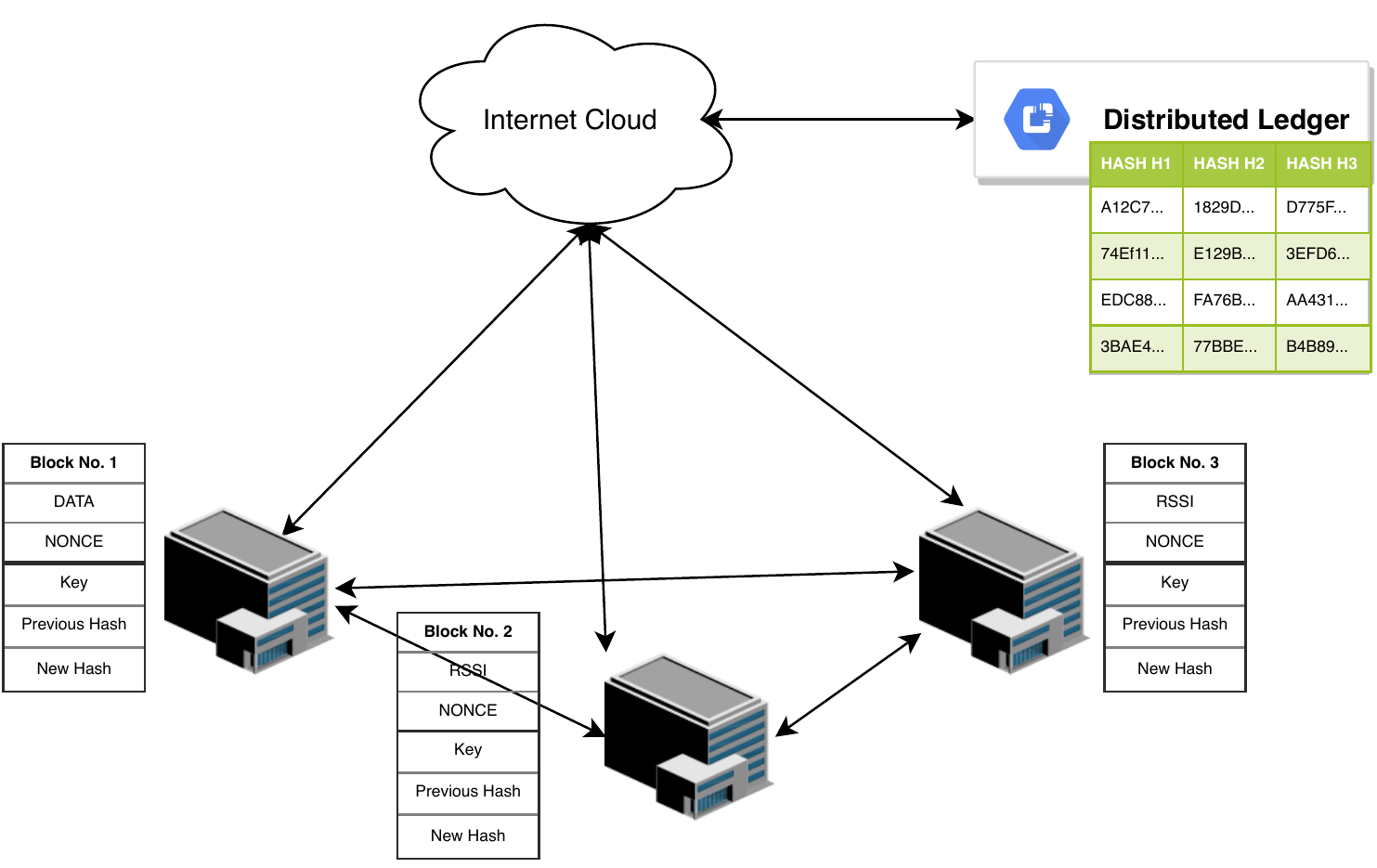}
\caption{Blockchain based connected healthcare units.}
\label{fig:blockchain_healtcare}
\end{figure}

Blockchain is the rapidly growing technology which became famous because of a virtual currency called Bitcoin. The use of blockchain is expanded to many fields \cite{ribeiro2020enhancing}. Blockchain enables privacy and security for data sharing \cite{celesti2020blockchain}\cite{alketbi2018blockchain}. A blockshain based IoT system presented in \cite{huh2017managing} stores the private key at IoT device while the public keys are stored at Etherum. Blockchain can be implemented for connected healthcare units as shown in Figure \ref{fig:blockchain_healtcare} in which all healthcare units are connected to each other. Each healthcare unit acts as a block and accurate data transfer is made possible by implementing blockchain based IoT network. For example, medical record of a patient received from one healthcare unit to another can be verified by generating a HASH and comparing it with all the HASH values present in the ledger.  

Blockchain technology can also be used to secure end to end data. The security and privacy preservation are made sure if IoT is integrated with blockchain. Important data of medical records and the record of all available health care kits and other resources are verified by any official by checking if the record is in its authentic form or is it forged \cite{celesti2020blockchain}. SHA1 algorithm is applied on the data along with private key ($K$) associated with medical health care unit, pseudo random nonce ($N$) and previous hash ($HASH_p$). 
Mathematically,
\begin{equation}
HASH = SHA1(Data, K_i, N_i, HASH_p)~. 
\end{equation}
All HASH values are sent to the cloud where a distributed ledger is created. The data can be checked in later date for its authentication. Even a single digit change in the report generates a different hash. Due to this, any forgery can be detected.

\begin{figure}[h]
\centering
\includegraphics[width=0.7\textwidth]{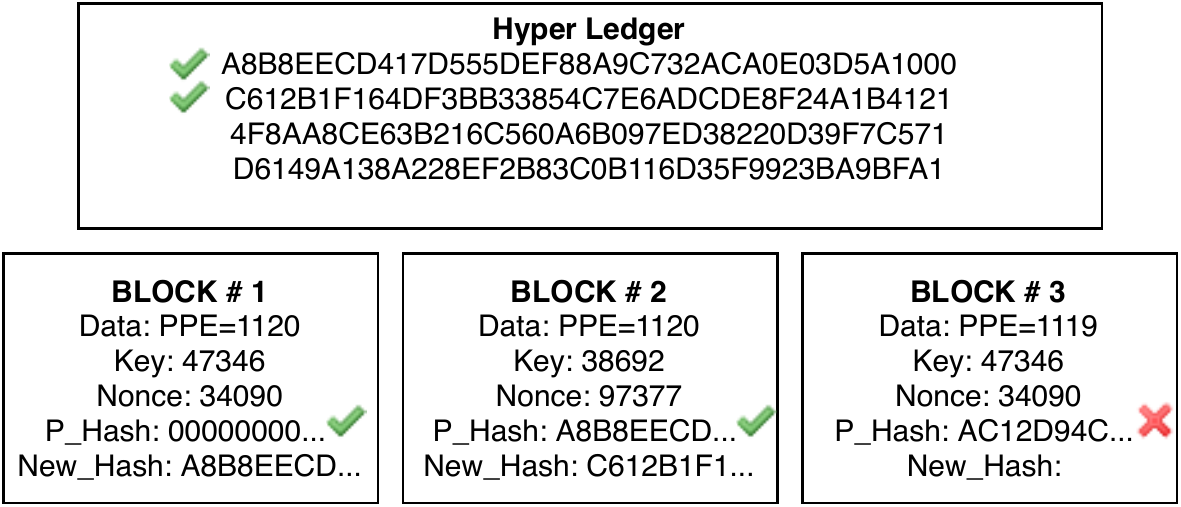}
\caption{Blockchian based security showing mismatch of hash in ledger for block \# 3.}
\label{fig:blockchain_ledger}
\end{figure}

This can be applied to the supply chain in which each supply point becomes a block and adds its hash to a decentralized ledger. The data (which could be the count of equipment) can be verified at any stage or precisely at the destination by looking at the HASH values in the ledger. If all HASH values match then the supply has reached successfully. Figure \ref{fig:blockchain_ledger} shows the same procedure in  which the ledger is updated with the HASH values generated by each block. At each block the HASH is varifiable while in the last block, the data is slightly changed and a different HASH is generated. While at the ledger, the HASH values generated and presented at the ledger is not matched with the HASH of block number 3.

\subsection{Cognitive Radio enabled IoT}
\begin{figure}
\centering
\includegraphics[width=0.8\textwidth]{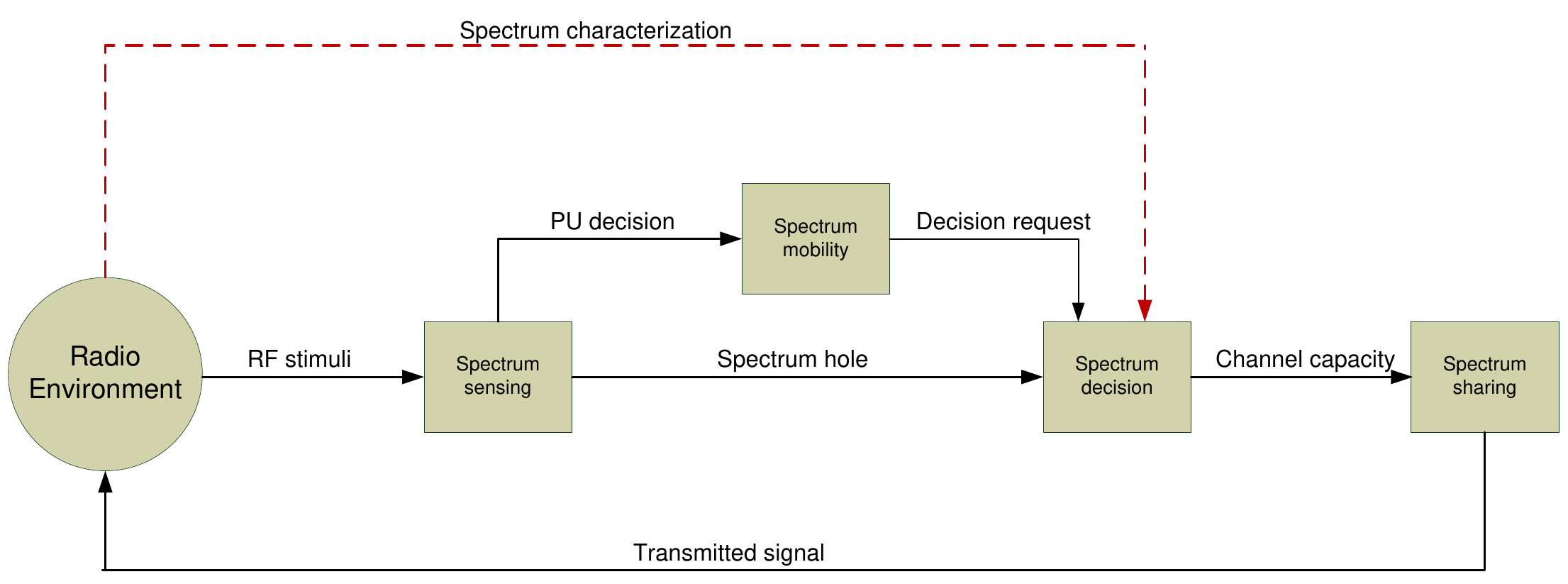}
\caption{The cognitive radio cycle.}
\label{fig:spectrum_sharing}
\end{figure}

Cognitive radio merged with IoT is called Cognitive Radio IoT (CRIoT) \cite{tarek2020new}. Spectrum allocation is always been done traditionally in a licensed fashion. It has been observed that most of the licensed spectrum is not completely utilized. Cognitive radios are proposed as a viable solution to the frequency reuse problem \cite{shakeel2019analysis}. While using cognitive radio parameters, IoT devices are capable of sensing the environment and adjust the configuration parameters automatically. The IoT devices sense the availability of free spectrum refereed as holes in the spectrum and communicate in the sensed holes without interfering with the licensed user called Primary User (PU) \cite{mazumder2020multi}. This helps in uninterrupted data communication and efficient utilization of licensed spectrum. Figure \ref{fig:spectrum_sharing} represents the whole cognitive radio cycle and is explained as under. 

\subsubsection{Spectrum sensing}
The process of detecting unused spectrum is known as spectrum sensing. Spectrum sensing is the basic feature of cognitive radio network. In order for IoT devices to communicate using licensed band, free holes in PU's spectrum need be detected. Therefore the IoT device monitors the available PU's spectrum and detect the available holes. After the spectrum has been sensed, the next step for IoT device is to select the most appropriate spectrum according to their QoS requirements. The spectrum decision is taken according to the radio environment and the statistical behavior of the PUs \cite{mazumder2020multi}. 
\subsubsection{Spectrum sharing}
The process to share the spectrum between PU and the IoT devices is referred as spectrum sharing. Spectrum sharing is an important step for reliable communication of licensed PUs and unlicensed IoT devices to share their spectrum with each other \cite{tarek2020new}. Since there may be more IoT devices to access the spectrum, spectrum sharing is very important in order to share the spectrum by preventing collision in overlapping portion of the spectrum. Spectrum sharing provides capabilities to share the spectrum resources opportunistically with multiple IoT devices which includes resource allocation to avoid interference caused to the PU. 
\subsubsection{Spectrum mobility}
The process of maintaining communication during spectrum transition is known as spectrum mobility. IoT device vacates the channel when a licensed user is detected. IoT device continues its transmission in another available portion of the spectrum. The transmission is switched to a new route or to a new band with minimum quality degradation \cite{yampurin2019frequency}.  

There are many methods to sense the availability or non availability of the PU in frequency spectrum. These methods are energy detection, matched filter (MF) detection, cyclosationary detection and feature detection. In later three methods, the system should know the prior knowledge about the primary signal while in energy detection method it is not required, hence making it more energy efficient \cite{sarala2020spectrum}. 


\subsection{Towards Millimeter wave for higher bandwidth}
With the advent of IoT, the demand of bandwidth has increased. For the organizations working on deploying IoT devices, the bandwidth shortage has motivated them in the exploration of the underutilized millimeter wave (mmWave) frequency spectrum for future IoT networks. mmWave ranges from 3 GHz to 300 GHz \cite{rappaport2013millimeter}. Spectrum at 28 GHz, 38 GHz, and 70-80 GHz looks especially promising for next-generation cellular systems. Because of large bandwidths, multi gigabits per second can be achieved. mmWave communication provides promising benefits in other application scenarios like wearable networks, vehicular communications, or autonomous robots etc. \cite{lv2018millimeter}.

As the frequency spectrum range is broad, more bandwidth is available at these frequencies. The capacity (C) is increased which solves the problem of scalability in IoT networks because mathematically,
\begin{equation}
C = BW \times log_2(1+SNR)~,
\end{equation}
where BW represents the bandwidth and SNR is the signal to noise ratio. Due to higher attenuation in free space, the same frequency is reused at shorter distances. The security and privacy are better because of limited range and narrow beam widths \cite{yilmaz2015use}. As the frequency is high then the wavelength is small and hence, small antenna size which helps in integrating the large array of antennas on a chip or Printed Circuit Boards (PCB). 

\subsection{Artificial Intelligence based IoT networks}
\begin{figure}[h]
\centering
\includegraphics[width=0.7\textwidth]{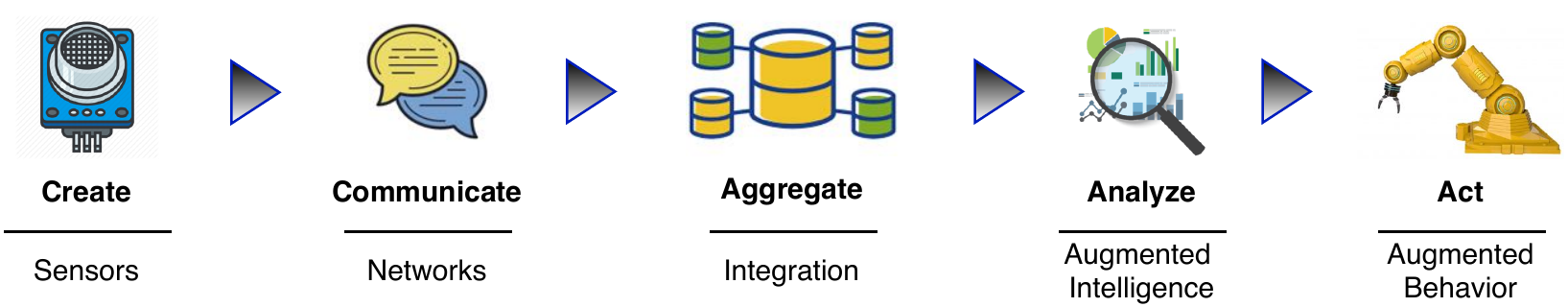}
\caption{Functional view of AI and IoT \cite{AI_article}.}
\label{fig:AI_functinal_view}
\end{figure}

There is no doubt in the fact that communication devices that are connected to one another over the Internet generate a lot of useful data. This data can be used to influence the decisions taken by the communication device. To make sense of this data, AI provides the context and thus providing extra information to assist the decision taken by a communication terminal. AI helps in detecting patterns and makes the terminal devices intelligent enough to learn from past patterns \cite{mahdavinejad2018machine}. This can happen in two ways:
\begin{enumerate}
\item \textbf{Predictive analysis:} The data is used to forecast what could be the outcome of a decision.
\item \textbf{Adaptive analysis:} What decisions can be taken based on past experiences to optimize the decision-making process?
\end{enumerate}
Machine learning, an AI technology, provides the ability to automatically identify patterns and detect anomalies in the data generated by smart sensors and devices. Machine learning approaches make operational predictions up to 20 times faster and more accurately. Other AI technologies, such as voice recognition and computer vision, help extract insights from data that required human confirmation \cite{AI_article}. The functional view of AI integrated with IoT is presented in Figure \ref{fig:AI_functinal_view}. 

Machine learning provides solutions in providing security for IoT networks, specifically in tackling the Denial of Service (DoS) attacks and replay attacks \cite{doshi2018machine}. Besides, it also provides solutions to resource efficient IoT networks when very small IoT devices are deployed having few Kilo Bytes of Random Access Memory (RAM) \cite{kumar2017resource}.

\section{Conclusion}
During the outbreak of the global challenge of COVID-19 pandemic, the reliance on technologies such as IoT, AI, Blockchain, Big Data Analytics, and Cloud computing has increased. IoT plays a major role in reducing the risks of coronavirus spread by providing platforms which helps in following the protocols defined by WHO. IoT based healthcare units provide timely response by medical staff to deal with COVID-19 patients. AI integrated with IoT helps in better forecasting the situation of future. IoT helps in providing support to academia and corporate professionals to carry out their tasks remotely. Blockchain based IoT networks helps in batter management of the supply chain and to detect any forgery in data. The challenges in implementing IoT networks can not be ignored. In order to deal with the scalability, CRIoT and mmWave based communication systems provide support to enable end-to-end communication. The need of lightweight security is also obvious because the IoT devices are small in size and large in number. The solution suggests to implement algorithms which have less computational cost. 

\subsection*{Acknowledgement} This  work  was  supported  by the Ministry of Health (MoH), Saudi Arabia grant No. 935. The authors, therefore, gratefully acknowledge the MoH technical and financial supports.

\bibliographystyle{unsrt}  
\bibliography{biblio}

\end{document}